\documentclass[prb,showpacs,twocolumn]{revtex4}
\usepackage{amsmath}
\usepackage{graphicx}

\newcount\minute
\newcount\hour
\newcount\hourMins
\def\now
{
  \minute=\time
  \hour=\time \divide \hour by 60
  \hourMins=\hour \multiply\hourMins by 60
  \advance\minute by -\hourMins
  \zeroPadTwo{\the\hour}:\zeroPadTwo{\the\minute}
}
\def\timestamp
{
  \today\ \now
}
\def\today
{
  \zeroPadTwo{\the\day}-\zeroPadTwo{\the\month}-\the\year%
}
\def\zeroPadTwo#1%
{
  \ifnum #1<10 0\fi
  #1%
}
\def \dif {\mathrm{d}}

\pacs{74.81.-g,75.70.-i,74.25.Dw,74.78.-w}
\date{\timestamp}

\begin{document}

\title{Superconducting film with randomly magnetized dots:
a realization of the 2D XY model with random
phase shifts}

\author{Zoran Ristivojevic}
\affiliation{Institut f\"{u}r Theoretische
Physik, Universit\"{a}t zu K\"{o}ln,
Z\"{u}lpicher Str. 77, 50937 K\"{o}ln,
Germany}
\affiliation{Materials Science Division, Argonne National Laboratory, Argonne, IL 60439, USA}

\begin{abstract}
We consider a thin superconducting film with
randomly magnetized dots on top of it. The dots
produce a disordered pinning potential for
vortices in the film. We show that for dots with
permanent and random magnetization normal or
parallel to the film surface, our system is
an experimental realization of the
two-dimensional XY model with random phase
shifts. The low-temperature superconducting
phase, that exists without magnetic dots,
survives in the presence of magnetic dots for
sufficiently small disorder.
\end{abstract}
\maketitle

\section{Introduction}

Since the pioneering papers by Berezinskii
\cite{Berezinskii+72} and Kosterlitz and Thouless
\cite{Kosterlitz+73} finite temperature phase
transitions in two dimensional (2D) systems,
which have a continuous symmetry specified by a
phase, have become a very active research field.
It turned out that two dimensional superfluids
\cite{Nelson+77} as well as thin superconducting
films
\cite{Beasley+79,Doniach+79,Turkevich79,Halperin+79}
have a Berezinskii--Kosterlitz--Thouless(BKT)
transition at a finite temperature $T_{2D}$. The
low-temperature phase is superfluid
(superconducting) and has quasi-long range order,
while the disordered high-temperature is normal
liquid (metallic) for superfluids
(superconductors). These systems are successfully
described by the 2D XY model, see Eq.~(\ref{H})
with $A_{ij}\equiv 0$. Theoretical predictions
were confirmed experimentally: a universal jump
in the superfluid density at $T_{2D}$
\cite{Bishop+78} and different current-voltage
characteristics of superconducting films below
and above $T_{2D}$ \cite{Resnick+81}.

\begin{figure}
\includegraphics[width=0.7\linewidth]{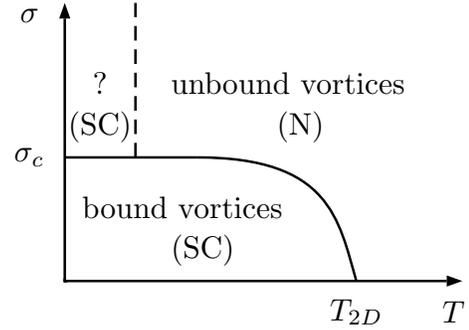}
\caption{Phase diagram of the 2D XY model with
random phase shifts. At low temeprature and weak
disorder $\sigma<\sigma_c$ a superconducting
phase (SC) occurs, and there vortices are bound in
pairs. Above $T_{2D}$ vortices are unbound and
the phase is non-superconducting (N). According
to some studies \cite{Holme+03,Yun+06,Chen+08} at
low temperatures and for strong enough disorder
appears another superconducting phase.}
\label{fig:phases}
\end{figure}

Introducing disorder through random phase shifts
in the 2D XY model physics becomes more
complex. The Hamiltonian of the 2D XY model
with random phase shifts reads
\cite{Rubinstein+83,Nattermann+95}
\begin{align}\label{H}
\mathcal{H}=-\frac{\varepsilon_0}{\pi}\sum_{i,j}
\cos(\phi_i-\phi_j-A_{ij}),
\end{align}
where the sum runs over all nearest neighboring
sites on a square lattice. $\phi_i$ denotes the
phase of the order parameter, while $A_{ij}$ are
quenched random phase shifts on bonds of the
square lattice produced by some kind of disorder.
We assume that $A_{ij}$ are Gaussian distributed,
uncorrelated, have the zero mean $\langle
A_{ij}\rangle=0$ and the variance $\langle
A_{ij}^2\rangle=\sigma$. The pure 2D XY model
contains two different kinds of excitations which
are decoupled. The spin-wave excitations describe
small changes of the phase $\phi$ and do not
drive any phase transitions. Another kind of
excitations are vortices, topological
excitations, and they are essential for the
existence of a BKT transition in the 2D XY model
at $T_{2D}$. Vortices are bound in pairs in the
low-temperature phase, and unbound at high
temperatures. Introducing the disorder in the form of
random phase shifts, the low-temperature
superconducting phase survives for weak enough
disorder, as has been shown first analytically
\cite{Nattermann+95,Cha+95,Jeon+95,Carpentier+98}
and then numerically
\cite{Maucourt+97,Kosterlitz+97}. The phase
diagram of the model (\ref{H}) is given in
Fig.~\ref{fig:phases}.

The Hamiltonian (\ref{H}) describes the
thermodynamic behavior of several disordered
systems, including two-dimensional ferromagnets with
random Dzyaloshinskii-Moriya interaction
\cite{Rubinstein+83}, Josephson junction arrays
with positional disorder \cite{Granato+86} and
vortex glasses \cite{Ebner+86}. In this paper we
will consider another system and show that it
also belongs to the class of systems that can be
described by the model (\ref{H}). It is a
ferromagnet-superconductor hybrid system. The
system consists of a thin superconducting film
covered by magnetic dots with permanent, but
random magnetization. For more details about
hybrid systems see recent reviews
\cite{Lyu+05,Vel+08}. We will show that our
system can be described by the Hamiltonian
(\ref{H}) under some conditions defined below.
Knowing the solution of the model (\ref{H}) we
can establish possible phases of the system.

Placed on top of the film, a single magnetic dot
with sufficiently large magnetization, normal to
the film surface, induces and pins vortices and
antivortices around it
\cite{Lyu+98,Sas+00,Mil+03,Ristivojevic08}. A
regular lattice of magnetic dots with constant
and equal magnetization is the source of a periodic
pinning potential for vortices in the film.
The magnetic dots induce periodic arrays of vortices
and antivortices in the film, provided the dot
magnetization is high enough
\cite{Milosevic+04,Priour+04}. The periodic
pinning is absent in the case of a random dot
magnetization, and there magnetic dots are a
source of disordered pinning potential for
vortices. Lyuksyutov and Pokrovsky
\cite{Lyu+98,Lyu+05} considered a superconducting
film with a magnetic dots array with random,
sufficiently strong magnetic moments and
concluded that the dot array induces the
resistive state in the film. They have not
considered the case when magnetic moments are
weak. In this paper we will show that the
superconducting state survives provided the
disorder is not too strong. Hybrid systems with a
regular lattice of magnetic dots with random
magnetization \cite{Villegas+08}, and without a
lattice, but homogeneously distributed dots on
top of a thin superconducting film \cite{Xing+08}
have been recently experimentally realized.

In the rest of the paper in section II we
introduce the model for our hybrid system and
give its solution by mapping it to the model
(\ref{H}). Section III contains discussions and
conclusions. Some technical details are postponed
to the appendix.

\section{Model and its solution}

We consider a thin superconducting film
characterized by the London penetration depth
$\lambda_L$, the coherence length $\xi$, a
thickness $d$ and a typical lateral dimension
$L$. The London penetration depth is in general
temperature dependent and we assume that our film
has the effective penetration depth
$\lambda=\lambda_L^2/d$ that exceeds film's
lateral dimension $L$. This limit is valid for
``dirty'' superconducting films
\cite{Beasley+79}. In addition, provided the bulk
critical temperature is larger than a critical
temperature $T_{2D}$ for vortex unbinding, the
film has a BKT transition at $T_{2D}$
\cite{Beasley+79,Doniach+79}.

We consider the dots with permanent random
magnetization placed on top of the film. They
produce a random potential $V$ for vortices in
the film. Assuming vortices of vorticities $n_i$
are places on a quadratic lattice with the lattice
constant $a$ (which is of the order of the
coherence length), the effective lattice
Hamiltonian for the system may be written as
\begin{align}\label{Hv}
\mathcal{H}_v=\sum_{i}\left[n_i^2 (E_c+U_v)+n_i
V_i\right]+\frac{1}{2}\sum_{i\neq j}n_i n_j
U_{vv}(\rho_{ij}),
\end{align}
where the sum runs over all lattice sites,
$\varepsilon_0=\phi_0^2/(16\pi^2\lambda)$, the
flux quantum is $\phi_0=hc/(2e)$, $V_i$ is the
random potential at site $i$, while $E_c$ is the
single vortex core energy which is of the order
$\varepsilon_0$. $U_v$ and $U_{vv}(\rho_{ij})$
are the single vortex energy and the interaction
energy of two vortices separated by a distance
$\rho_{ij}$ respectively, and for $L<\lambda$
read \cite{Pea65,Chaikin+}
\begin{align}\label{U}
U_v=\varepsilon_0\ln\frac{L}{a},\quad
U_{vv}(\rho_{ij})=2\varepsilon_0\ln\frac{L}{\rho_{ij}}.
\end{align}
Using the expressions (\ref{U}) it is useful to
rewrite the Hamiltonian (\ref{Hv}) in the form
\begin{align}\label{Hvnew}
\mathcal{H}_v=\sum_i\left(n_i^2 E_c+n_i
V_i\right)-\varepsilon_0\sum_{i\neq j}n_i
n_j\ln\frac{\rho_{ij}}{a}+N^2 U_v,
\end{align}
where we have introduced the total vorticity of
the system $N=\sum_i n_i$. In the limit $L\gg a$
and without the disorder potential $V_i=0$, the last term in Eq.~(\ref{Hvnew})
penalizes the total energy for nonzero total
vorticities \cite{Blatter+94,Minnhagen81}, so one has $N=0$. Then, a superconducting film described by the
model (\ref{Hvnew}) has a BKT transition at
the temperature $T_{2D}=\varepsilon_0(T_{2D})/2$,
where $\varepsilon_0(T_{2D})$ denoted that one
should take value $\lambda(T_{2D})$ renormalized
by the presence of vortices
\cite{Kosterlitz+73,Beasley+79,Doniach+79}.

Next, we consider effects of the random
potential. First we consider the case when the dots
have random magnetization parallel to the film
surface. We model them as magnetic
dipoles placed on top of the film at lattice
sites. To characterize statistical properties of
the dots, we assume that the $x-$ and $y-$ components of
the magnetic moment at site $i$ are Gaussian
distributed, have zero mean value and are
uncorrelated from site to site:
\begin{align}\label{m:correlator}
\langle m_{i\alpha}\rangle=0,\quad \langle
m_{i\alpha}
m_{j\beta}\rangle=\mathcal{M}^2\delta_{ij}\delta_{\alpha\beta},
\quad\alpha,\beta=x,y
\end{align}
where $\langle\ldots\rangle$ denotes an average
over disorder. Since the dots have random magnetic
moments, $\mathcal{M}$ is the
measure for a typical magnetic moment of a
magnetic dot at some lattice site.

The interaction energy between a single dot having the
magnetic moment $\mathbf{m}$ parallel to the film
surface and a vortex of vorticity $n$ placed at a
relative distance $\boldsymbol{\rho}$ from the
dot can be calculated using the approach
developed in Ref.~\cite{Erd+02}, and reads
\cite{Ristivojevic08}
\begin{align}\label{Umvperp}
U_{mv}^{\parallel}(n,\rho)&=\frac{\mathbf{m}\cdot\boldsymbol{\rho}}{\rho}
\frac{n\phi_0}{2\pi} \int_0^\infty\dif
k\frac{kJ_1(k\rho)}{1+2\lambda k}\\\notag
&=\frac{\mathbf{m}\cdot\boldsymbol{\rho}}{\rho}
\frac{n\phi_0}{16\lambda^2}\left[H_1\left(\frac{\rho}{2\lambda}\right)
-Y_1\left(\frac{\rho}{2\lambda}\right)-\frac{2}{\pi}\right],
\end{align}
where $J_1$ is the Bessel function of the first
kind, $H_1$ is the Struve function and $Y_1$ is
the Bessel function of the second kind
\cite{Abramowitz}. Notice that the interaction
energy between the magnetic dipole and the vortex
can be simply written in the form
$-\mathbf{m}\cdot \mathbf{B}_v$, where
$\mathbf{B}_v$ is the magnetic field produced by
the vortex at the dipole position
\cite{Carneiro04,Ristivojevic08}. In the limit
$\rho\ll\lambda$ the interaction energy
(\ref{Umvperp}) reads
\begin{align}
U_{mv}^{\parallel}(n,\rho)=\frac{\mathbf{m}\cdot\boldsymbol{\rho}}{\rho}
\frac{n\phi_0}{4\pi\lambda\rho}.
\end{align}
The random site potential $V_i$ is given as a sum
over the lattice
\begin{align}\label{Vi}
V_i=\sum_{j\neq
i}\frac{U_{mv}^{\parallel}(n_j,\rho_{ij})}{n_j}
&=\frac{\phi_0}{4\pi\lambda}\sum_{j\neq i}
\frac{\mathbf{m}_j\cdot\boldsymbol{\rho}_{ij}}{\rho_{ij}^2}.
\end{align}
By summing over $j\neq i$ in the previous
expression we avoid the short scale cutoff
divergence of the interaction energy
(\ref{Umvperp}) at $\rho=0$, which exists because
the dots are placed at the top of the film in our
model. In reality magnetic dipoles are separated
from the film surface by some small distance
$\sim a$. Since $V_i$ is a sum of many
independent random variables it is Gaussian
distributed (notice here that the assumption
about Gaussian distribution for $m_{i\alpha}$ is
not necessary condition for $V_i$ to be Gaussian
distributed; it is sufficient that dots have a
distribution with a finite variance). Its mean,
variance and site to site correlations can be
calculated and read
\begin{align}\label{V:mean}
&\langle V_i\rangle=0,\\
\label{V:variance} &\langle
V_i^2\rangle=2\pi\frac{\mathcal{M}^2\varepsilon_0}{\lambda
a^2}
\ln\frac{L}{a}+\mathcal{O}(1),\\
\label{V:sitetosite} &\langle(V_i-V_j)^2\rangle=
4\pi\frac{\mathcal{M}^2\varepsilon_0}{\lambda
a^2} \ln\frac{\rho_{ij}}{a} +\mathcal{O}(1),
\end{align}

The model (\ref{Hvnew}) with the disorder
potential (\ref{Vi}) which has properties
(\ref{V:mean}), (\ref{V:variance}) and
(\ref{V:sitetosite}) matches the vortex
part of the 2D XY model with random phase shifts
\cite{Rubinstein+83,Tang96}. From the solution of
the model (\ref{H})
\cite{Nattermann+95,Tang96,Carpentier+98} we know
its phase diagram, see Fig.~\ref{fig:phases}. At
zero temperature there is a critical value for
the typical magnetic moment per unit length
\begin{align}\label{Mcritical}
\frac{\mathcal{M}_c}{a}=\frac{\phi_0}{8\pi\sqrt{2\pi}}.
\end{align}
For $\mathcal{M}<\mathcal{M}_c$ the system has no free vortices $N=0$,
while for $\mathcal{M}>\mathcal{M}_c$ the disorder spontaneously creates and
induces unbound vortices, and one generally has $N\neq 0$. A simple argument \cite{Korshunov+96} based on Eq.~(\ref{Hvnew}) which compares the energy loss for the single vortex creation and the energy gain due to disorder fluctuations also gives the critical value (\ref{Mcritical}) for zero temperature. The connection between $\sigma$ introduced as the disorder strength in
Eq.~(\ref{H}) and the typical magnetic moment
$\mathcal{M}$ introduced in
Eq.~(\ref{m:correlator}) is
$\sigma=\left(\frac{4\pi^2 \mathcal{M}}{\phi_0
a}\right)^2$, while the random phase is
$\mathbf{A}_i=\frac{4\pi^2}{\phi_0
a}\mathbf{e}_z\times\mathbf{m}_i$ where the $x-$
($y-$)component of $\mathbf{A}_i$ corresponds to
the disorder $A_{ij}$ on horizontal (vertical)
bond at site $i$.

In the case that the dot magnetization is normal
to the film surface, similar to
Eq.~(\ref{Umvperp}), for the interaction energy
between a dot of magnetization $m$ and a vortex
of vorticity $n$ separated by a distance $\rho$
we get
\begin{align}\label{Umvnorm}
U_{mv}^{\perp}(n,\rho)&=m \frac{n\phi_0}{2\pi}
\int_0^\infty\dif k\frac{kJ_0(k\rho)}{1+2\lambda
k}\\\notag &=m
\frac{n\phi_0}{16\lambda^2}\left[Y_0\left(\frac{\rho}{2\lambda}\right)
-H_0\left(\frac{\rho}{2\lambda}\right)-\frac{4\lambda}{\pi\rho}\right].
\end{align}
The previous expression simplifies for
$\rho\ll\lambda$ and reads
\begin{align}
U_{mv}^{\perp}(n,\rho)=m\frac{n\phi_0}{4\pi\lambda\rho}.
\end{align}
By assuming that the film is covered by magnetic
dots with random magnetization normal to the film
surface (along the $\hat{\mathbf{z}}$ direction), satisfying
\begin{align}
\langle m_{iz}\rangle=0,\quad \langle m_{iz}
m_{jz}\rangle=\mathcal{M}^2\delta_{ij},
\end{align}
for the site random potential we get
\begin{align}\label{Viprim}
V_i'=\sum_{j\neq
i}\frac{U_{mv}^{\perp}(n_j,\rho_{ij})}{n_j}=
\frac{\phi_0}{4\pi\lambda}\sum_{j\neq i}
\frac{m_{jz}}{\rho_{ij}},
\end{align}
which, as we show in the appendix, is equivalent
to $V_i$ and hence has statistical properties
(\ref{V:mean})-(\ref{V:sitetosite}). We may
conclude that the system with magnetic moments
normal to the film has the same phase diagram as
in the case of moments parallel to the film
surface.

\section{Discussions and conclusions}


Having shown the equivalence between our system
and the vortex part of the 2D XY
model with random phase shifts, we may infer some
properties of the former. The phase diagram is
given in Fig.~\ref{fig:phases}. The low
temperature and low disorder phase is
superconducting. There vortices and antivortices
are bound in pairs. The current--voltage
characteristic for $T\to T_{2D}$ and $T<T_{2D}$
and for weak disorder is expected to be very
similar to the one of Halperin and Nelson
\cite{Halperin+79} for the pure case, $V\sim
I^3$, possibly with a small correction due to the
disorder. This phase has zero linear resistivity.
The high temperature phase is metallic and has a
nonzero linear resistivity. There free vortices
dissipate energy and produce the linear
current--voltage characteristic $V\sim I$.

In Ref.~\cite{Lyu+98} the conclusion about
the resistive state of the film when the dots
with normal magnetization are present relies on
the assumption that the randomly magnetized dots
pin vortices of vorticity $\pm 1$. These pinned
vortices serve as a source of the random
potential for other bound vortices, which unbind
and fill deep valleys of the random potential.
These unbound vortices lead to the resistive
state of the film. We agree that this scenario
occurs for sufficiently strong disorder when the
dots can induce and pin vortices. A single dot
can induce and pin quite different configurations
of vortices and antivortices regarding its
magnetic moment\cite{Ristivojevic08}. We expect that a random
lattice of dots can also pin, from site to site,
quite a different number of vortices and
antivortices which produce different potential
than one assumed in Ref.~\cite{Lyu+98}.
However, our conclusion, that the resistive state
in films occurs when the disorder is sufficiently
strong, agrees with the one from
Ref.~\cite{Lyu+98}. Moreover, we give the
strength of the disorder above which the
resistive state occurs.

By making a comparison between $\mathcal{M}_c$ and the
magnetic moment $m_{1c}$ of a single magnetic dot
with normal magnetization necessary to induce and
pin an extra vortex in the film, we obtain
$m_{1c}\approx\mathcal{M}_c\sqrt{8\pi}(b/a)\ln(L/a)$,
where $b$ is the distance between the dipole and
the film surface. In addition, knowing that the
value $\mathcal{M}_c$ corresponds not to typical but rare
magnetic moments from the tail of distribution
\cite{Nattermann+95,Korshunov+96}, we may
conclude that even a very rare magnetic dot in
the film that has the magnetic moment $\mathcal{M}_c$ is
not able to induce and pin vortices. Such pinned
vortices served as a source of random potential
in Ref.~\cite{Lyu+98}.


In this paper we have considered the dots as magnetic
dipoles. This fact is unimportant as
long as the dot size is not too big with respect
to the lattice constant. What is crucial for any
kind of magnetic dots is their interaction with a
vortex which decays as $1/\rho$, which is
universal for any geometrical shape of dots, when
the vortex is sufficiently far from the dot. The
magnetic field produced by a vortex decays as
$1/\rho$ and the interaction energy dot--vortex
universally decays, regardless of the shape of
the dot. This form of the interaction produces
logarithmically diverging, with the system size,
variance of the disorder potential that is
characteristic for the Hamiltonian (\ref{H}).

Recently, the question of a possible third phase
for strong disorder and at low temperatures has
been raised in numerical studies of the model
(\ref{H}) with uniformly random phase $A_{ij}$ in
the interval $[-r\pi,r\pi]$ with $0\le r\le 1$,
with conflicting results about its existence
\cite{Holme+03,Katzgraber03,Chen+08}. While for
$r<r_c\approx 0.37$ \cite{Kosterlitz+97} it is
accepted that the superconducting phase survives
at low temperature, the case $r>r_c$ is still
under debate. An experimental study of Josephson
junction arrays with positional disorder
\cite{Yun+06} supports the existence the third
phase. The new phase is superconducting according
to the experiments of Ref.~\cite{Yun+06} and
numerical investigations of Ref.~\cite{Holme+03}.
In the limit $\sigma\to\infty$ the model
(\ref{H}) and the so-called gauge glass limit,
$r=1$, are equivalent, and we expect that the
conjectures about the phase diagram for the case
of uniformly distributed phase apply as well to
the Gaussian distribution of the phase. The
existence of the third phase is a challenging
question that could be experimentally resolved by
using magneto-superconducting hybrid systems as a
realization for the model (\ref{H}).

To conclude, we have shown that a thin
superconducting film covered by magnetic dots
with random magnetization provides an
experimental realization for the two-dimensional
XY model with random phase shifts. The phase
diagram of the latter model helped us to conclude
that a low-temperature superconducting phase of a
superconducting film without dots survives when
the dots are placed on top of the film, provided
their magnetization is not too large.

\section*{ACKNOWLEDGMENTS}

This work is financially supported by the DFG
under the grant NA222/5-2 and through SFB 608. The
author thanks Prof.~T.~Nattermann and Prof.~V.~Pokrovsky for discussions and A.~Petkovi\'{c} for
reading the manuscript and helpful suggestions.

\section{APPENDIX}

In this appendix we prove that expressions for
the disorder potential produced by magnetic moments
paralel to the film surface (\ref{Vi}) and normal
to the film surface (\ref{Viprim}) are
equivalent. Rewriting the expression
$\mathbf{m}_j\cdot\boldsymbol{\rho}_{ij}/\rho_{ij}^2$
from Eq.~(\ref{Vi}) as
$m_j\cos(\alpha_j-\alpha_{\rho})/\rho_{ij}$,
where $\cos\alpha_j=\mathbf{m}_j\cdot
\mathbf{e}_x/m_j$ and
$\cos\alpha_\rho=\boldsymbol{\rho}_{ij}\cdot
\mathbf{e}_x/\rho_{ij}$, we will in the following
show that the distribution of random variable
$m_{rj}=m_j\cos(\alpha_j-\alpha_\rho)$ (which is
the projection of $\mathbf{m}_j$ onto
$\boldsymbol{\rho}_{ij}$) is Gaussian,
with zero mean and the variance $\mathcal{M}^2$.

By assumption (\ref{m:correlator}), the components
of the magnetic moment $m_{jx}$ and $m_{jy}$ are
Gaussian distributed and have the distribution
function
\begin{align}
p(t)=\frac{1}{\sqrt{2\pi}\mathcal{M}}
\exp\left(-\frac{t^2}{2\mathcal{M}^2}\right)\quad
t=m_{jx},m_{jy}.
\end{align}
Then the distribution function of the magnetic
moment $m_j=\sqrt{m_{jx}^2+m_{jy}^2}$ is
\begin{align}
p(m_j)=\int_{-\infty}^{\infty}\dif m_{jx}
\int_{-\infty}^{\infty}\dif m_{jy}
p(m_{jx})p(m_{jy})\notag\\ \times\delta\left(m_j-\sqrt{m_{jx}^2+m_{jy}^2}\right)
=\frac{m_j}{\mathcal{M}^2}
\exp\left(-\frac{m_j^2}{2\mathcal{M}^2}\right),
\end{align}
while the angle $\alpha_j$ between $m_{jx}$ and
$m_j$ is uniformly distributed in the interval
$[0,2\pi)$ and has the distribution
$p(\alpha_j)=1/(2\pi)$. The distribution of the
random variable $m_{rj}$ is
\begin{align}
p(m_{rj})=\int_0^{\infty}\dif m_j\int_0^{2\pi}\dif\alpha_j
p(m_j)p(\alpha_j)\notag\\ \times\delta\left(m_{rj}-m_j\cos(\alpha_j-\alpha_{\rho})\right).
\end{align}
The previous expression can be most easily
evaluated first by taking the Fourier transform
of $p(m_{rj})$
\begin{align}
\tilde{p}_k=\int_0^\infty\dif
m_j\int_0^{2\pi}\dif\alpha_j\frac{m_j}{2\pi\mathcal{M}^2}
\exp\left(-\frac{m_j^2}{2\mathcal{M}^2}\right)\notag\\ \times\exp[i
k m_j\cos(\alpha_j-\alpha_{\rho})]=
\exp\left(-\frac{k^2\mathcal{M}^2}{2}\right),
\end{align}
and then taking the inverse Fourier transform of
the previous expression. It leads to
\begin{align}
p(m_{rj})=\frac{1}{\sqrt{2\pi}\mathcal{M}}
\exp\left(-\frac{m_{rj}^2}{2\mathcal{M}^2}\right).
\end{align}

In that way we have proved that the random
potential (\ref{Vi})
\begin{align}
V_i=\frac{\phi_0}{4\pi\lambda}\sum_{j\neq i}
\frac{\mathbf{m}_j\cdot\boldsymbol{\rho}_{ij}}{\rho_{ij}^2}=
\frac{\phi_0}{4\pi\lambda}\sum_{j\neq i}
\frac{m_{rj}}{\rho_{ij}}
\end{align}
matches the random potential (\ref{Viprim}). We conclude that frozen magnetic dipoles parallel to
the film create the same random potential for
vortices in the film as magnetic dipoles normal
to the film, provided both are Gaussian
distributed.


\end{document}